\documentclass[reprint,twocolumn,superscriptaddress,aps,pre]{revtex4-1}

\usepackage{amsmath}
\usepackage{color}
\usepackage{graphicx}
\usepackage{epstopdf}
\usepackage{mathtools}
\usepackage{texdraw}
\usepackage{tikz}
\usepackage{transparent}
\usepackage{textcomp}
\usepackage{gensymb}
\usepackage{graphicx}
\usepackage[separate-uncertainty=true]{siunitx}
\usepackage{placeins}
\usepackage{lipsum}
\usepackage{verbatim}
\usepackage{xcolor}
\usepackage{subfigure}
\usepackage{bbold}
\usepackage{xifthen}
\usepackage{MnSymbol}

\newcommand{\be}{\begin{equation}}
\newcommand{\ee}{\end{equation}}
\newcommand{\bea}{\begin{align}}
\newcommand{\eea}{\end{align}}

\newcommand{\Nb}{N_\mathrm{b}}
\newcommand{\Ns}{N_\mathrm{s}}
\newcommand{\Nzero}{N_0}
\newcommand{\Nss}{N_\mathrm{sss}}
\newcommand{\tss}[1]{t_{\mathrm{ss},#1}}
\newcommand{\elleq}[1][]{%
\ifthenelse{\equal{#1}{}}{\ell_\mathrm{eq}}{\ell_{\mathrm{eq},#1}}%
}
\newcommand{\dli}{\delta \ell_i}
\newcommand{\calE}{\mathcal{E}}
\newcommand{\calC}{\mathcal{C}}
\begin{document}
\author{Anwesha Bose}
\email{a.bose@tue.nl}
\affiliation{Department of Applied Physics, Eindhoven University of Technology, Den Dolech 2, 5600MB Eindhoven, The Netherlands}
\author{Mathijs F.J. Vermeulen}
\affiliation{Department of Applied Physics, Eindhoven University of Technology, Den Dolech 2, 5600MB Eindhoven, The Netherlands}
\author{Cornelis Storm}
\affiliation{Department of Applied Physics, Eindhoven University of Technology, Den Dolech 2, 5600MB Eindhoven, The Netherlands}
\affiliation{Institute for Complex Molecular Systems, Eindhoven University of Technology, Den Dolech 2, 5600MB Eindhoven, The Netherlands}
\author{Wouter G. Ellenbroek}
\email{w.g.ellenbroek@tue.nl}
\affiliation{Department of Applied Physics, Eindhoven University of Technology, Den Dolech 2, 5600MB Eindhoven, The Netherlands}
\affiliation{Institute for Complex Molecular Systems, Eindhoven University of Technology, Den Dolech 2, 5600MB Eindhoven, The Netherlands}

\title{Self-stresses control stiffness and stability in overconstrained disordered networks}
\date{\today}
\begin{abstract}
\noindent We investigate the interplay between pre-stress and mechanical properties in random elastic networks. To do this in a controlled fashion, we introduce an algorithm for creating random freestanding frames that support exactly one state of self stress. By multiplying all the bond tensions in this state of self stress by the same number---which with the appropriate normalization corresponds to the physical pre-stress inside the frame---we systematically evaluate the linear mechanical response of the frame as a function of pre-stress. After proving that the mechanical moduli of affinely deforming frames are rigourously independent of pre-stress, we turn to non-affinely deforming frames. In such frames, pre-stress has a profound effect on linear response: not only can it change the values of the linear modulus---an effect we demonstrate to be related to a suppressive effect of pre-stress on non-affinity---but pre-stresses also generically trigger bistable mechanical response. Thus, pre-stress can be leveraged to both augment the mechanical response of network architectures on the fly, and to actuate finite deformations. These control modalities may be of use in the design of both novel responsive materials and soft actuators. 
\end{abstract}
\maketitle

\section{Introduction}
The concept of mechanical metamaterials \cite{bertoldi2017flexible, zadpoor2016mechanical,reis2015designer}---structures that inherit purposely targeted, non-standard mechanical response to stress or strain from a particular spatial architecture, rather than from intrinsic properties of the materials they are composed of---has gained massive traction over the past decade. At macroscopic length scales, research into mechanical metamaterials has helped achieve, in systematic fashion, desirable properties such as auxeticity \cite{evans2000auxetic,prawoto2012seeing, reid2018auxetic, rocks2017designing, babaee20133d}, chirality \cite{frenzel2017three,huang2016design}, (origami-like) deployability and actuation \cite{schenk2013geometry, overvelde2016three, lv2014origami, silverberg2014using,coulais2018multi, coulais2016combinatorial} and anomalously high strength in ultra-lightweight lattices \cite{zheng2014ultralight,schaedler2011ultralight}. The central and profound insight that has enabled these breakthroughs is, that the mechanical response of a generalized material is due to combination of {\em (i)} the mechanical properties of its constituent(s), {\em (ii)} its spatial architecture (i.e., its void distribution) and {\em (iii)} its mechanical preconditioning; the configuration of internal stresses that resides on the spatial architecture. Each of these three factors may be targeted in design, but historically only {\em (i)} has been explored. Much of the current work in metamaterials can be understood as the exploring of a design space that stretches out along directions {\em (ii)} and {\em (iii)}.

These concepts are simultaneously, and in parallel, finding their way into microscopic, molecular scale designs for polymeric matter where, likewise, they are allowing access to unusual mechanical response that is difficult to attain in pure bulk matter. Hydrogels, in particular, have proven to be a terrific canvass for exploring directions {\em (ii)} and {\em (iii)}; harnessing residual stresses, spatial composition and non-covalent binding have yielded materials that, through purely mechanical and geometrical effects, exhibit greatly enhanced strength and toughness, structural adaptivity and recyclability \cite{Gong2003,Gong2010,Ducrot2014,Tsukeshiba2005,Huang2007,singh2011physical,yang2012prestressed}.

While the utility and successes of these microscopic design approaches in soft materials are irrefutable, a crucial difference between macroscopic and hydrogel metamaterials remains. Macroscopic materials are meticulously organized in space (by direct design of the entire structure or a unit cell), and can be loaded at will. Smart hydrogel architectures, in contrast, are microscopically disordered. It is, and will likely remain, impossible to place the polymers at the well-specified positions and orientations typical of macroscopic metamaterials. Yet, even these disordered polymeric materials show similarly responsive properties. This raises the questions which of the macroscopic design strategies---material choice, spatial architecture and mechanical preconditioning---may be implemented in disordered soft materials, and how the anomalous mechanical response prevails in spite of the disorder.

In this paper, we address these questions for control modality {\em (iii)}: mechanical preconditioning in disordered elastic networks. Our approach is rooted in macroscopic tensegrities: architectures that are geometrically overconstrained (i.e., that possess fewer degrees of freedom than they have constraints) and as a result have one or more so-called states of self stress (SSS)~\cite{Calladine1978}. These SSS play a key role in determining mechanical response; previous work in physics \cite{Wilhelm2003, huisman2011internal, lubensky2015phonons,paulose2015topological,Vermeulen2017,Sussman2016} and mechanics \cite{connelly1996second, guest2006stiffness, guest2011stiffness, schenk2007zero} links their existence to the bulk rigidity of spring networks and granular packings. Even in networks in which the actual self stress is zero, the mere knowledge of which SSS the network geometrically allows can be used to compute elastic moduli~\cite{wyartthesis,lubensky2015phonons}. But what happens to these moduli, and other mechanical properties, when the actual stresses are nonzero? Starting from a network with only one single SSS, a continuum of geometrically indistinguishable network architectures can be defined, differing only in the tension configuration along their bonds. The key question now is whether, and if so how, the geometrically identical networks in such a family of self-stressed states \emph{do} differ mechanically. What, in short, is the effect of engaging the SSS in overconstrained networks?

Our paper is laid out as follows. First, we recall the general framework of geometrical mechanics of frames. Then, we detail a method to generate disordered frames with exactly one SSS and no floppy modes. This permits the cleanest discussion of our central question, but may be generalized to frames possessing multiple SSS. We describe, in general terms, the effect of SSS on the non-affinity and along the way note that, within this model, self-stresses can not augment mechanical moduli via affine deformation modes. Our disordered SSS frames generally do not deform affinely, and the remainder of our paper details two main effects of engaged pre-stresses on self-stressed disordered frames: they change the moduli, and at sufficiently high values can destabilize frames to produce various types of bistability.

\section{Geometrical mechanics: Maxwell-Calladine counting and states of self stress in finite frames}\label{Sec1}

Following, largely, the conventions of \cite{lubensky2015phonons} we define a {\em frame} to be a spatial distribution of $\Ns$ pointlike nodes, connected by $\Nb$ bonds. Bonds are either on the boundary, or they are internal to the frame. Any initial configuration of the frame in $d$ dimensions is geometrically completely specified by a length-$d \Ns$ vector ${\bf \sf X}_0=\left\{\vec r_{0,i}\right\}_{i=1}^{\Ns}$ containing all node positions. These initial node positions define the \emph{initial lengths} $\ell_{0,k}$ of the bonds---if bond $k$ connects nodes $i$ and $j$ then $\ell_{0,k}=|\vec r_{0,j}-\vec r_{0,i}|$, and we may collect all these initial lengths into a length-$\Nb$ vector  ${\bf \sf L}_0=\left\{ \ell_{0,k}\right\}_{k=1}^{\Nb}$. Deformed states may now be defined in reference to these initial configurations  by specifying the vector of node displacements ${\bf \sf \delta X}=\left\{\delta \vec r_i\right\}_{i=1}^{\Ns}$. With any deformation $\mathsf{\delta X}$ comes a set of bond length changes ${\bf \delta \sf L}=\left\{ \delta \ell_{k}\right\}_{k=1}^{\Nb}$. To linear order in the bond displacements, ${\bf \sf \delta X}$ and ${\bf \sf \delta L}$ are related through the $\Nb \times d \Ns$ {\em compatibility matrix} ${\cal Q}^T$:
\be\label{Qteq}
{\bf \sf \delta L}={\cal Q}^T{\bf \sf \delta X}\, .
\ee
In general, each bond $k$ in the frame carries a tension $t_k$, directed (for central force networks) along the bond unit vector $\hat n_k$. The tension is a signed scalar quantity, and in the following we adopt the convention that a positive value of $t_k$ corresponds to a tensile force in bond $k$. That is, if $t_k$ is positive then bond $k$ pulls the two nodes it connects towards the middle of bond $k$. Similar to above, we may collect all bond tensions into a single length-$\Nb$ vector ${\bf \sf T}=\left\{ t_{k}\right\}_{k=1}^{\Nb}$. The tensioned bonds exert forces on the nodes that they connect, that may be computed as the vector sum of all the forces in these bonds. Again, these forces constitute a length-$d\Ns$ vector ${\bf \sf F}=\left\{ \vec f_{i}\right\}_{i=1}^{\Ns}$. Node forces and the bond tensions inhabit the same geometry that relates node positions and bond extensions, and therefore are related to each other in similar fashion:
\be\label{Qeq}
{\bf \sf F}=-{\cal Q}{\bf \sf T}\, ,
\ee
with ${\cal Q}$ the $d \Ns \times \Nb$ {\em equilibrium matrix}, the transpose of the compatibility matrix. For a finite frame to be at mechanical equilibrium, the net force on each of its nodes must be zero: ${\bf \sf F}=0$. Now, the null space of ${\cal Q}^T$ contains those node displacements ${\bf \delta \sf X}$ that do not result in any change in any of the bond lengths; ${\bf \delta \sf L}=0$. These displacements (which, in general, involve multiple or even all nodes moving in concerted fashion) are called the {\em zero modes} of the frame. All 2D frames have at least 3 zero modes; the trivial 3 correspond to 2 translations and a single rotation. Zero modes other than these three are called {\em floppy modes} in the physics literature, and {\em mechanisms} in the mechanical literature. They represent zero energy deformations of the frame. The dimension of the null space of ${\cal Q}^T$ is thus the number of independent zero modes.

The null space of ${\cal Q}$ contains those tensions ${\bf \sf T}$ that do not result in any net force on any of the nodes; ${\bf \sf F}=0$. Of course, these force balance equations are always trivially solved by ${\bf \sf T}=0$, but we will be interested in the nontrivial solutions. Such configurations of nonzero bond tensions, which still give rise to overall mechanical equilibrium are called {\em states of self-stress} (SSS). They may arise for purely topological reasons, or because of special geometries that affect the rank of $\mathcal{Q}$ such as crystalline order or strain-induced rearrangements~\cite{Calladine1978,thorpe95,During2014,Vermeulen2017}.

The original Maxwell counting argument \cite{Maxwell} asserts that the number of zero modes of the frame equals the number of degrees of freedom (here, in $d$ dimensions) minus the number of constraints imposed by the springs (one per bond);

\be
N_0=d N_s-N_b\, .
\ee

However, since every state of self stress represents a redundant connection in the system, the number of SSS must be subtracted from the number of bonds which leads to the modified, Calladine-Maxwell count
\be
N_0=d N_s-(N_b-N_{ss})\, .
\ee
Calling $\nu=\dim \ker {\cal Q}^T-\dim \ker {\cal Q}$ the {\em index} of ${\cal Q}$ (and noting that, as explained above, $\nu=N_0-N_{ss}$) we obtain the general Calladine-Maxwell 'index theorem' \cite{lubensky2015phonons}

\be
\label{eq:maxwellcalladine}
\nu=d N_s-N_b\, .
\ee 
We call a frame {\em rigid} when it has no zero modes other than the three trivial ones; $N_0=3$.

Focusing for a moment on the SSS, let us suppose that a given frame has a single state of self stress; that is, the dimension of the solution space of the equation $\mathcal{Q}\,\mathsf{T}\ = 0$ is one. We are then free to choose a basis vector $\mathsf{T_{\rm ss}}=(\tss{1},\tss{2},\ldots,\tss{\Nb})$ for this Kernel space in any way we like. As soon as we do, a continuous family of possible self-stresses may be defined through a scalar parameter $\alpha$---obviously any vector $\alpha \mathsf{T_{\rm ss}}$ is also an admissible, equilibrium SSS of the exact same frame but at a different self-stress. This degeneracy is at the heart of our present paper: we will ask how the mechanical properties change as we change the self-stress in geometrically identical frames. To fix the normalization, we choose our reference basis vector such that that the actual bond tensions in the network may be written as $t_k=\Pi \mathsf{T}_\mathrm{ss,k}$, where $\Pi=-\frac12\mathsf{Tr}(\sigma)$ (with $\sigma$ the Cauchy stress tensor) is the pressure carried by the boundary bonds of the frame. In practice, this means that we normalize $\mathsf{T_{\rm ss}}$ such that the sum of the four boundary tensions satisfies $\tss{1}+\tss{2}+\tss{3}+\tss{4}=-2$. This choice also facilitates comparisons between freestanding frames and osmotically swollen periodic frames with the same internal bond geometry and bond forces.  

So far, we have discussed the mechanics of frames in entirely geometrical terms. Generally, an elastic ``constitutive relation'' relates the bond extensions and the bond tensions. In most of the following, we will assume that the tension of a single bond, stretched to a length $\ell_k$ and with a rest length $\elleq[k]$ is given by
\be\label{Hooke}
t_k=\frac{Y}{\elleq[k]}\left(\ell_k-\elleq[k]\right)\, ,
\ee
with $Y$ a force scale that, if one imagines constructing the frame out of beams made from a linearly elastic material, can be interpreted as the the Young's modulus of that material times the cross-sectional area of the beams. Note the sign convention, a stretched bond has a positive $t_k$ and pulls its nodes towards its middle. The above implies that the elastic energy $\varepsilon_k$ of a single stretched bond is
\be
\varepsilon_k=\frac{Y}{2\elleq[k]}(\ell_k-\elleq[k])^2\, .
\ee
Summing over all bonds and using Eq.~(\ref{Qteq}), this gives the total elastic energy of the frame deformed relative to a force-balanced reference configuration, in terms of the node displacements $\mathsf{\delta X}$. For small $\mathsf{\delta X}$, this may be written (summing over repeated indices) as
\be\label{en0}
\mathcal{E}(\delta\mathsf{X})=\mathcal{E}(0)+\frac12 {\bf \delta \sf X}_i ({\cal H})_{ij} {\bf \delta \sf X}_j\, .
\ee 
The elements of the stiffness matrix or Hessian ${\cal H}$ are then defined by ${\cal H}_{ij} \equiv \frac{\partial^2 {\cal E}}{\partial {\bf \delta \sf X}_i\partial {\bf \delta \sf X}_j}$ and encode the rigidity of the frame against general deformations. In particular, negative eigenvalues of the Hessian matrix signal mechanical instabilities, since the corresponding eigenvector defines a direction of decreasing energy in Eq.~(\ref{en0}); deformations in that direction happen spontaneously. For stress-free initial configurations, $\mathcal{H}$ can be written as $\mathcal{H}=\mathcal{QKQ}^T$~\cite{wyartthesis}, with $\mathcal{K}$ a diagonal matrix encoding the stiffnesses $Y/\elleq[k]$, but we will use it in its full form, so that we can use Eq.~(\ref{en0}) to establish stability boundaries in pre-stressed frames.

\section{Self-Stressed Frames: Generation and Mechanical Analysis}\label{Sec2}
\begin{figure}[h]
	\includegraphics[width=\linewidth]{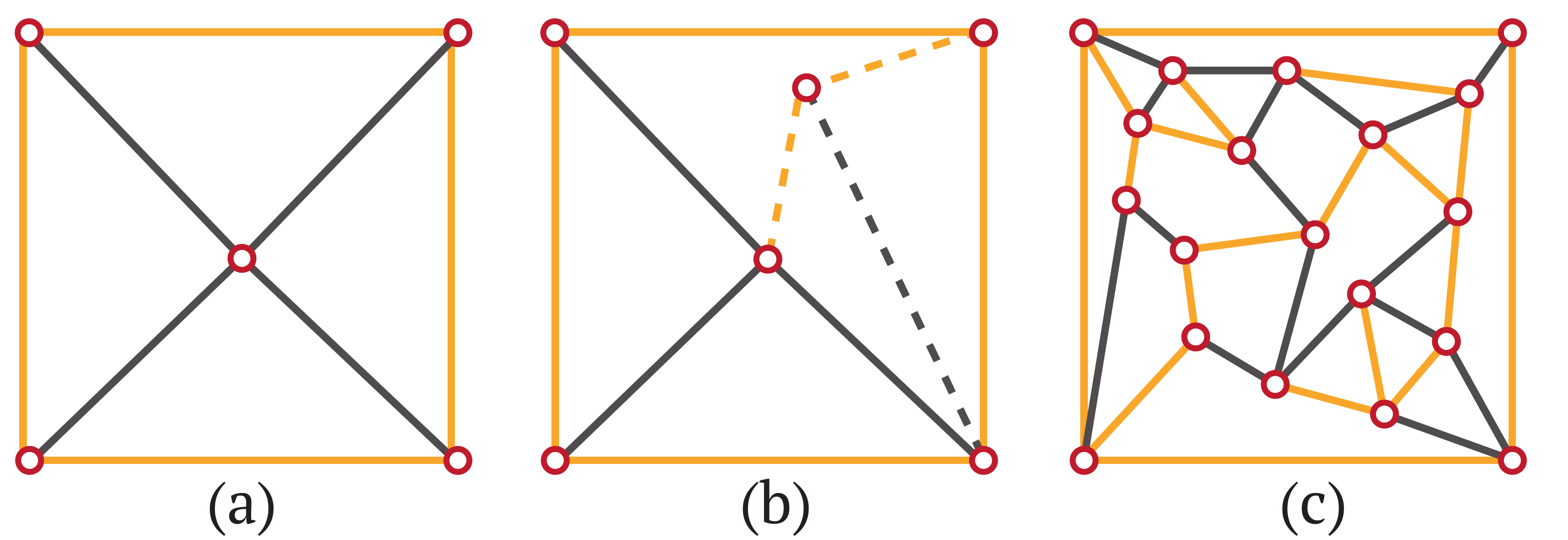}
	\caption{(a) A simple network with single SS state, (b) Edge splitting moves on frame (c) Anisotropic frame network with compressed (yellow) and tensile (gray) bonds under free-standing conditions. The stressed bonds represent the state of self-stress.}
	\label{fig:frame}
\end{figure}
We model inhomogeneously stressed materials using spring networks with explicit boundaries, which we call \emph{freestanding frames}. In absence of external
pre-loading, the total stress in these frames is zero, unlike periodic systems where the fixed shape and size of the periodic unit cell can induce pre-stress. Self-stresses are therefore the only pre-stress in these systems, and for general networks the actual self-stress is some nonzero linear combination of the states of self-stress (SSSs) of the system. To keep things tractable, we focus on frames with a single SSS, ranging from Snelson's X, the canonical square with $\Ns=5$ nodes connected by $\Nb=8$ bonds shown in Fig.~\ref{fig:frame}(a)~\cite{Snelson1965}, to more complex networks such as that shown in Fig.~\ref{fig:frame}(c). The entire family of freestanding frames is constructed such that they do not have any floppy modes and they obey $2\Ns-\Nb=2$, so that according to Eq.~(\ref{eq:maxwellcalladine}) they are guaranteed to have exactly one state of self-stress.

The procedure to generate the more complex frames such that they are still guaranteed to possess exactly one state of self-stress is a repeated application of a modification known as the \emph{bond splitting} Henneberg moves~\cite{henneberg1911}. As illustrated in Fig.~\ref{fig:frame}b, the move amounts to adding a node in an existing bond and connecting this node to another existing node. At the same time, the geometry of the network is kept as generic as possible by displacing the newly added node such that the two halves of the bond it split are no longer collinear. This procedure trivially conserves the difference  $\Nzero - \Nss$ because it adds two degrees of freedom as well as two constraints. Furthermore, as long as the three bonds of the newly created node all point in different directions, $\Nzero$ and $\Nss$ are individually conserved. In order to keep the distribution of bond lengths narrow, we iteratively apply this procedure to the longest bond in the network. Fig.~\ref{fig:frame}c shows an example of a result of this procedure after 14 iterations. By construction, this frame has one state of self-stress and no floppy modes. The coordination number of the frame after $n$ iterations is $z=2\Nb/\Ns=(16+4n)/(5+n)$: The average number of bonds per node increases with each iteration and asymptotically approaches the isostatic limit $z=4$. These finite frames serve as a convenient template to conduct our numerical calculations to study the mechanical response.

To isolate the effect of the pre-stress on the mechanical response, we analyze our frames as spring networks with a fixed geometry, varying the rest lengths $\elleq$ of the springs in order to tune the amount of pre-stress, parametrized by the boundary-bond pressure $\Pi$. The fixed geometry is specified via the positions of the nodes, which set the \emph{initial lengths} $\ell_0$ of the springs. Now, the initial tensions in the frame are no longer zero and Hooke's law (Eq.~\ref{Hooke}) gives
\begin{equation}
\label{eq:setelleq}
\Pi\,\tss{i} = \frac{Y}{\elleq[i]}(\ell_{0,i} - \elleq[i])~,
\end{equation}
from which, for each spring $i$, the rest lengths $\elleq[i]$ can be determined as a function of pre-stress $\Pi$. In the rest of the paper, we fix the arbitrary overall force scale by setting $Y=1$. Requiring that Eq.~(\ref{eq:setelleq}) has a positive solution for $\elleq[i]$ implies a constraint $\Pi\tss{i}>-1$, which gives a lower (upper) bound on $\Pi$ via bonds for which $\tss{i}$ is positive (negative). The equilibrium length of bond $i$ will diverge as these bounds are approached. We restrict our analysis of these frames to values of $\Pi$ that fall within the physically accessible range of pre-stresses
\be
\Pi_{\rm min}<\Pi<\Pi_{\rm max}
\ee
with
\be\label{minmax}
\Pi_{\rm min}\!=\!\frac{-1}{\max_i(\tss{i}|\tss{i}>0)}\, , \Pi_{\rm max}\!=\!\frac{-1}{\min_i(\tss{i}|\tss{i}<0)}\,.
\ee
The total energy stored in the frame now depends on the magnitude of the pre-stress $\Pi$, via the extension of the springs in the undeformed state $\ell_{0,i}-\elleq[i]$, and on any subsequent deformations which change the lengths of the springs further by an amount $\dli$. Summing the harmonic spring energy over all springs, we obtain
\begin{equation}
\label{eq:MainEnergy}
\calE(\Pi) = \frac12 \sum_{i=1}^{N_b} \left(\frac{1}{\elleq[i]}\right)\,(\ell_{0,i} + \dli - \elleq[i])^2~.
\end{equation}
Thus, $\calE(\Pi)$ represents both the work needed to prepare the pre-stressed frame as well as
the work involved in deforming the pre-stressed frame. Rewriting this energy in such a way that it references only the initial state as specified by the initial lengths $\ell_{0,i}$ and the pre-stress $\Pi$ using Eq.~(\ref{eq:setelleq}), we write
\begin{eqnarray}\label{threeterms}
\calE(\Pi) &=&\frac12 \sum_{i=1}^{N_b}(1+\Pi\,\tss{i})\,\ell_{0,i}\left(\frac{\Pi \tss{i}}{1+\Pi \tss{i}}+\frac{\dli}{\ell_{0,i}}\right)^2\nonumber\\
&\equiv& \calC(\Pi)+\calE(0)+\Delta\calE(\Pi)\,.
\end{eqnarray}
Here, 
\be
\calC(\Pi)=\frac12 \sum_{i=1}^{\Nb} \frac{\Pi^2\,\tss{i}^2}{1+\Pi\,\tss{i}}\,\ell_{0,i}~,
\ee
is the constant which measures the work performed on the frame to bring it to the pre-stressed state (i.e., the shift in the zero-strain energy),
\be
\calE(0)=\frac12 \sum_{i=1}^{N_b}\frac{\dli^2}{\ell_{0,i}} 
\ee
is the deformation energy at zero pre-stress, and 
\begin{align}\label{eq:mainEnergyDiff}
\Delta\calE(\Pi) = \sum_{i=1}^{\Nb} \Pi \tss{i}\left( 1+ \frac{\dli}{2\,\ell_{0,i}}\right)\,\dli~
\end{align}
is an additional, new pre-stress dependent change in the deformation energy. We will be investigating under what conditions a nonzero $\Pi$ changes the mechanical moduli of the frame, and therefore will consider what happens to each of the three components of the energy when the boundary of the frame is subjected to a deformation. That is, nodes $\vec r_B$ on the frame boundary are displaced to new positions $\vec r_B'$ according to some deformation gradient tensor $\Lambda$:
\be
\vec r_B'=\Lambda\cdot \vec r_B\, .
\ee
The interior (non-boundary) nodes in a frame subject to such a deformation will, in general, respond non-affinely exploiting their freedom to displace differently from boundary nodes to reduce the incurred bond stretching energy. For general nodes, therefore, the displacement may be written as
\be
\vec r_i'=\Lambda\cdot \vec r_i+\vec\delta_i\,.
\ee
With this, the displacement vector $\vec u_k$ of bond $k$, connecting nodes $i$ and $j$, 
\be
\vec u_k=(\vec r_j'-\vec r_i')-(\vec r_j-\vec r_i')\, ,
\ee
may be expressed in terms of $\Lambda$ and $\hat n_k$, the unit vector along bond $k$, as
\be
\vec u_k=\ell_{0,k}(\Lambda-\mathbb{1}_2)\cdot \hat n_k+\vec \Delta_k\,;
\ee
$\vec \Delta_k=\vec \delta_j-\vec \delta_i$ is the non-affine part of the bond displacement vector.
\begin{figure}
\includegraphics[width=\linewidth]{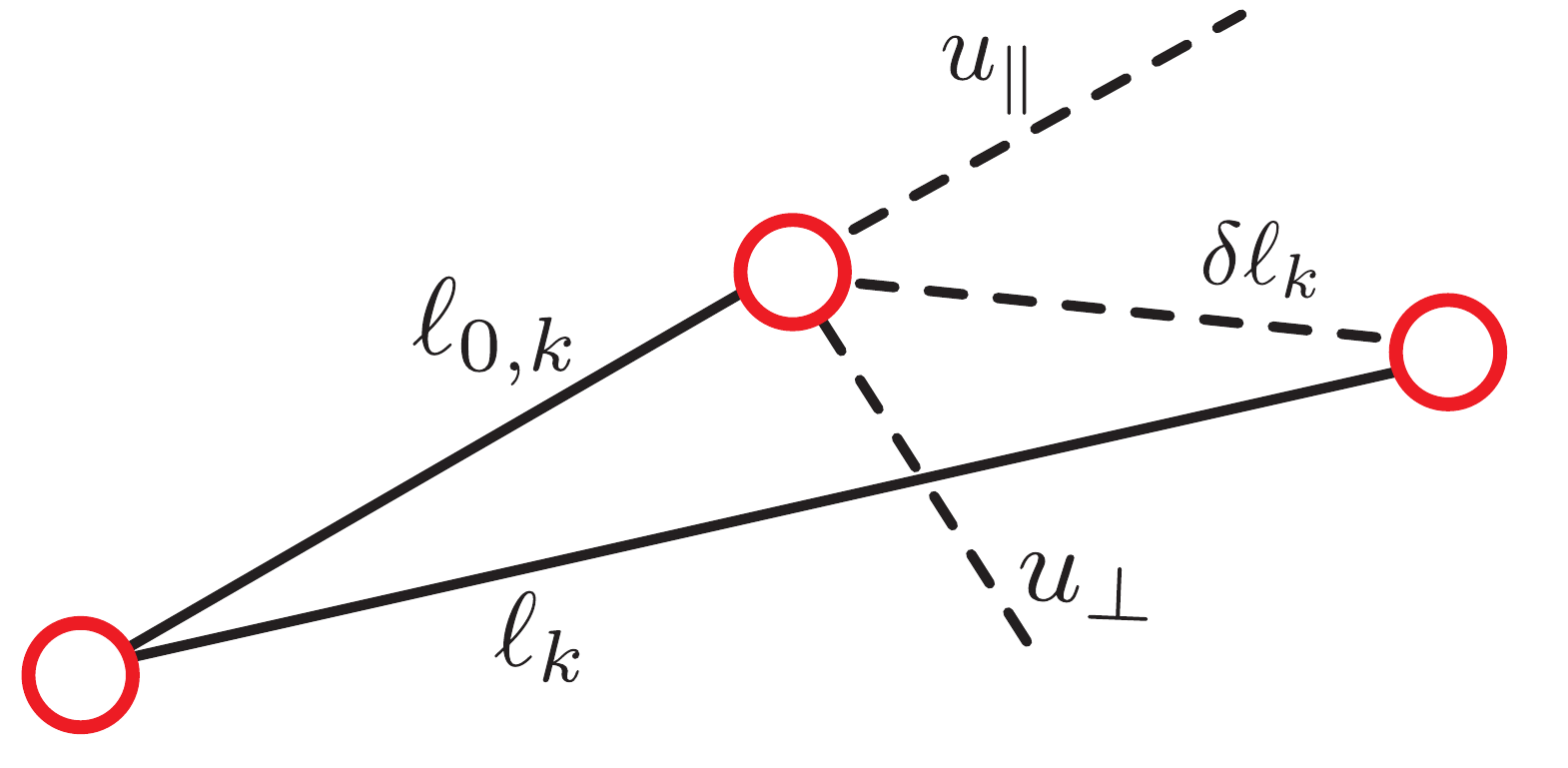}
\caption{Decomposing the relative displacement of nodes $i$ and $j$ into components parallel ($u_{\parallel}$) and perpendicular ($u_{\perp}$) to the original bond vector
provides a convenient way to express the change in length $\delta\ell_k=\ell_k-\ell_{0,k}$.}
\label{fig:uperp}
\end{figure}
Now, the change in the length of bond $k$ can be written in terms of the parallel ($u_{k,\parallel}=\hat n_k\cdot \vec u_k$) and perpendicular ($u_{k,\perp}=\hat p_k\cdot \vec u_k$ with $\hat p_k$ a unit vector perpendicular to $\hat n_k$) components of the relative displacement of the two nodes (see Fig.~\ref{fig:uperp}), as
\begin{align}
\delta \ell_k &= |(\vec r_{0,j}-\vec r_{0,i})+(\delta \vec r_{j}-\delta \vec r_{i})|-\ell_{0,k} \nonumber \\
&= |(\ell_{0,k}+u_{k,\parallel})\hat n_{k}+u_{k,\perp}\hat p_{k}|-\ell_{0,k} \nonumber \\
&= \ell_{0,k}\biggl(1+2\ell_{0,k}^{-1}u_{k,\parallel}+\ell_{0,k}^{-2}(u_{k,\parallel}^2+u_{k,\perp}^2) \biggr)^{1/2}-\ell_{0,k} \nonumber
\end{align}
To access the moduli, we must now ask how each of the three terms in the energy of the pre-stressed frame, Eq.~(\ref{threeterms}), is affected by the deformation $\Lambda$. The first term, $\calC(\Pi)$, is unaffected as it does not depend on the deformation. The second term, $\calE(0)$ does depend on the deformation but not directly on the pre-stress (we come back to this in a moment). The third term encodes the direct coupling between pre-stress and deformation and, using Eq.~(\ref{eq:mainEnergyDiff}), reduces to
\begin{multline}
\label{eq:ediffmultiline}
\Delta\calE(\Pi)=\sum_{k=1}^{\Nb}\Pi\tss{k} u_{k,\parallel} \, + \\ \frac12 \sum_{k=1}^{\Nb}\left(\frac{\Pi\tss{k}}{\ell_{0,k}}\right)(u_{k,\parallel}^2+u_{k,\perp}^2)~.
\end{multline}
Note, that Eq.~(\ref{eq:ediffmultiline}) is \emph{not} an expansion in $u$. Due to the specific structure of Eq.~(\ref{eq:mainEnergyDiff}), the square roots drop out and there are no terms beyond those quadratic in $u$.

A logical place to start the analysis is to consider an affine deformation; one where $\vec \Delta_k=0$ for all bonds. In that case, every point $\vec{r}$ in the reference frame is mapped onto its deformed image $\vec{r}'$ according to just the deformation gradient tensor $\Lambda$: $\vec{r}' = \Lambda \cdot \vec{r}$. In appendix~\ref{sec:Appendix}, we show that in this particular case something curious happens: both displacement-dependent terms in Eq.~(\ref{eq:ediffmultiline}) become proportional to the overall stress tensor $\sigma_{\alpha\beta}$. In particular, the linear term can be written as
\be
\label{eq:linear}
\sum_{k=1}^{N_b} t_{ss,k} \, u_{k,\parallel} = (\Lambda -\mathbb{1}_2)_{\alpha \beta} \cdot \sigma_{\alpha \beta}~,
\ee
and the quadratic term as
\be
\label{eq:quad}
\sum_{k=1}^{N_b}\left(\frac{{t}_{ss,k}}{\ell_{0,k}}\right)(u_{k,\parallel}^2 + \, u_{k,\perp}^2) = 
\left( \Lambda-\mathbb{1}_2 \right)_{\mu \alpha}\left( \Lambda-\mathbb{1}_2 \right)_{\mu \beta} \cdot \sigma_{\alpha \beta}~.
\ee
In free-standing frames, the total stress tensor $\sigma_{\alpha \beta}$ is zero, because there are no external forces. This brings us to a first conclusion about the effect of pre-stress: in an affine system,
\be
\calE(\Pi)=\calC(\Pi)+\calE(0)\, ,
\ee
and since the energy shift $\calC(\Pi)$ does not depend on the deformation, all derivatives with respect to the deformation including those that yield the moduli are unchanged. In short, engaging the state of self stress by applying a pre-stress to the frame can never change the mechanical moduli of an affinely deforming frame, in the case where the extensional moduli of the bond springs vary inversely with bond length. While these conditions may appear restrictive, we note that the class of systems to which they apply includes all crystalline frames with one node per unit cell. In such systems, nodes are constrained by local symmetries to move affinely. For the large class of these systems that also have a single bond length, the scaling of the spring constant with the equilibrium length is immaterial. All such systems are forbidden from having pre-stress dependent moduli. We suggestively attribute the fact that this surprising result has not been reported before to the usual focus on periodic systems, rather than our free-standing frames. We will address the connection between our findings and the tensegrity literature in the disucssion.

Fortunately, there is also a large class of non-affinely deforming frames to which the above restriction does not apply. Do these have pre-stress dependent moduli? The answer to this question is yes, and to see how it arises we now assume that $\vec\Delta_k\neq 0$ for some, or all, of the bonds of the frame. Doing so again leaves $\calC(\Pi)$ unaffected, as it does not depend on the deformation. The other two terms {\em do} change. We may split the total energy into an affine ($A$) part (i.e., independent of $\vec\Delta_k$) and two new non-affine ($NA$) parts as
\be\label{NAEn}
\calE(\Pi)=\calE^A(\Pi)+\calE^{NA}(0)+\Delta\calE^{NA}(\Pi)\,
\ee
where we recognize the part of the non-affine deformation energy correction that does not explicitly depend on the pre-stress,
\begin{eqnarray}\label{nazero}
\calE^{NA}(0)&=&\frac12\sum_{k=1}^{N_b}\ell_{0,k}\left[\biggl(|\Lambda\cdot \hat n_k|^2+\Gamma_k\right)^{1/2}\!-\!1\biggr]^2\nonumber \\&&\qquad -\frac12\sum_{k=1}^{N_b}\ell_{0,k}\biggl[|\Lambda\cdot \hat n_k|-1\biggr]^2\, ,
\end{eqnarray}
and another part that couples the pre-stress and the non-affinity
\be\label{NAcoup}
\Delta\calE^{NA}(\Pi)=\frac12 \sum_{k=1}^{N_b}\ell_{0,k}\Pi \tss{k}\Gamma_k\,.
\ee
Notation has been condensed somewhat by introducing
\be
\Gamma_k\equiv\frac{2}{\ell_{0,k}}(\Lambda\cdot \hat n_k)\cdot \vec \Delta_k+\frac{1}{\ell_{0,k}^2}|\vec \Delta_k|^2\, .
\ee
Clearly, the affine limit $\vec \Delta_k=0$ corresponds to $\Gamma_k=0$ in which limit both NA energy terms are zero. While it might appear that $\calE^{NA}(0)$ does not depend on the pre-stress since $\Pi$ does not appear, in fact it does. The reason for this is that the ground state energy of the non-affinely deforming frame is defined as the {\em minimum} of Eq.~(\ref{NAEn}) over the non-affine node displacements $\vec \delta_i$. Through the coupling term Eq.~(\ref{NAcoup}), this minimum will be attained at different values of $\Gamma_k$, which in turn affects the value of the non-affine deformation energy Eq.~(\ref{nazero}).

While the mathematical conditions for this minimization may be written down analytically, due to the combination of vectorial (terms proportional to the orientation of $\vec\Delta_k$, particularly in relation to the unit bond vector $\hat n_k$) and scalar (terms proportional to $|\vec\Delta_k|$) aspects, solving them in the general case is, in practice, possible only numerically. A further complication is, that the SSS components $\tss{k}$ may be both positive and negative, and indeed that both signs generally occur within a single SSS. As a result, there is no straightforward way to read off the high- or low pre-stress limits. In the special case, however, where all internal pre-stresses are extensional (that is, all $\tss{k}$ for internal bonds are positive) we conclude, that the non-affine energy terms favor $\vec \Delta_k$'s that are small and/or perpendicular to the bond unit vector (both of which contribute to the minimization of $\Gamma_k$ which dominates the non-affinity-pre-stress coupling term in the energy (Eq.~(\ref{NAcoup})) at high $\Pi$). Based on this, we can conclude that for purely extensional SSS's, large pre-stresses will suppress non-affinity, which should manifest in increasingly small values of $|\vec\Delta_k|^2$ at higher $\Pi$.

In general, and as we will detail in the following, pre-stresses due to engaged states of self stress can and do affect the moduli of non-affinely deforming frames. Since the affine deformation energy represents an upper bound to the actual (non-affine) energy (affinity is a constraint, the release of which will lead the system down, not up, in the energy landscape), the linear modulus of a generic frame must be lower than that of the same frame deforming affinely. In the following, we show that changing the pre-stress $\Pi$ allows for tuning of the moduli up to their affine values, and that the mechanism by which this happens is the suppression of non-affine displacements by the pre-stress. 

Having shown that an affinely deforming frame is insensitive to the magnitude of pre-stresses in the network we now analyze systems with non-affine deformations. We characterize these using the differential (linearized) shear modulus. Our main results are ({\em i})  that the differential shear modulus \emph{does} depend on the pre-stress, that ({\em ii}) this happens via a generally repressive coupling between the non-affine displacements and the pre-stress, and that ({\em iii}) pre-stress can induce instabilities in the frame, a fact most readily observed by noting there are values of $\Pi$ at which the shear modulus becomes negative. In Section \ref{Sec3}, we detail findings ({\em i}) and ({\em ii}), turning to finding ({\em iii}) in Section \ref{Sec4}.

\section{Linear response of self-stressed frames: Modulus and non-affinity}\label{Sec3}
\begin{figure}
\includegraphics[width=\columnwidth]{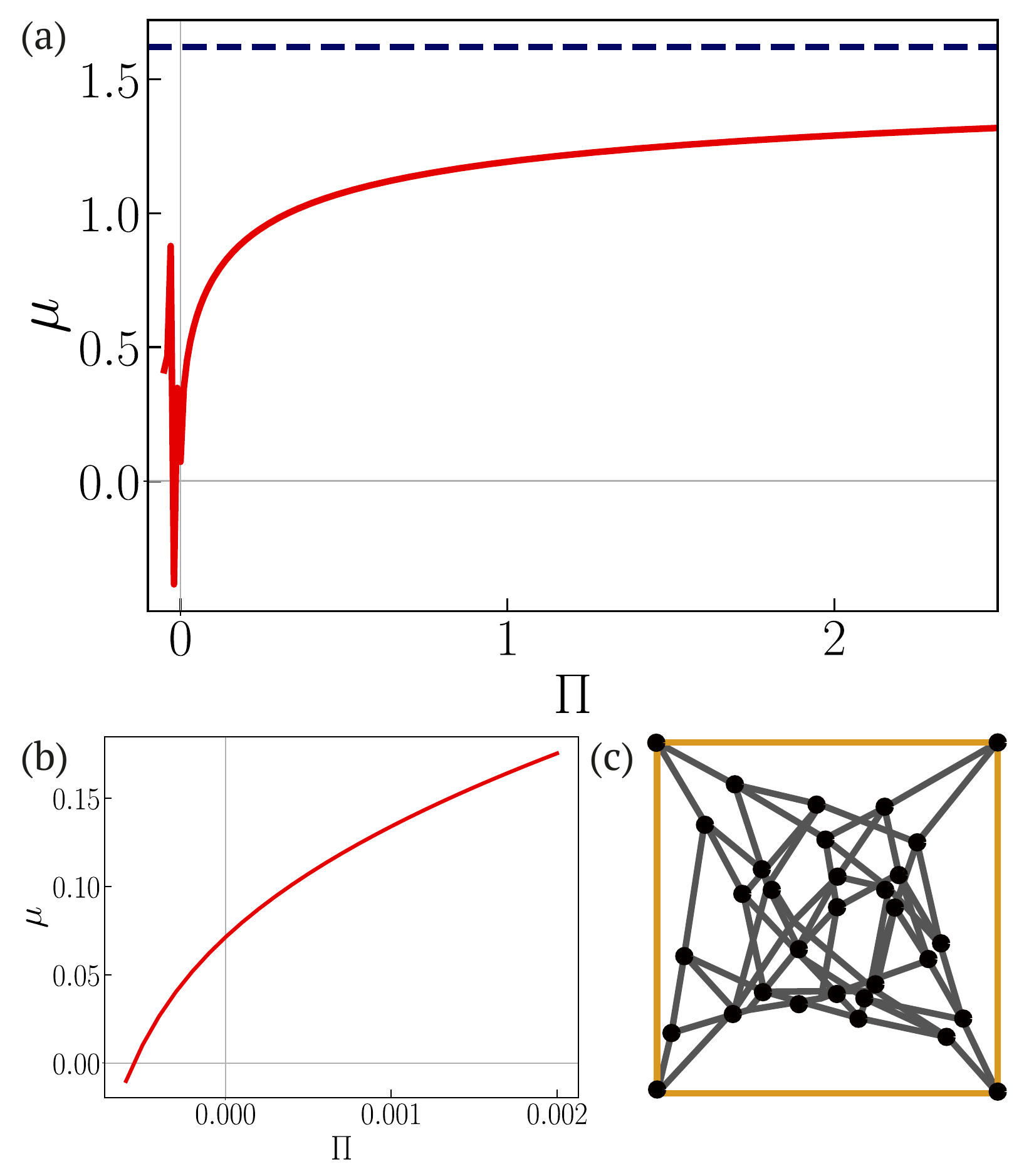}
\caption{(a)~Pre-stress dependence of the shear modulus $\mu$ of a non-affinely deforming free-standing frame (shown in (c)). As the self-stress $\Pi$ is increased the shear modulus (red) of geometrically indistinguishable frames changes, and is bounded by the affine shear modulus $\mu_{\rm aff}$ (dashed blue line). For this particular frame $\Pi_{\rm min}=-1.82$,  $\Pi_{\rm max}=2.52$ and $\mu_{\rm aff}=$1.62. (b)~Zoom-in of the region near $\Pi=0$, showing that the modulus becomes negative for $\Pi < 0$, signalling that the frame becomes unstable to simple shear. (c)~Frame with 64 tensile (gray) and 4 compressive (yellow) bonds connecting across 35 nodes.}
\label{fig:affineMod}
\end{figure}

Our procedure for analyzing response, both linear and nonlinear, is the following. First, we generate a frame applying iterated bond-splitting Henneberg moves to Snelson's X as described in Section {\ref{Sec2}. Then, we dial in a value for pre-stress $\Pi$; this leaves the geometry of the frame unaltered but is reflected in a set of bond equilibrium lengths $\elleq[i]$. In those cases where we want frames with exclusively tensile forces on the internal bonds, we temporarily set the rest length to zero and allow the internal bonds to relax keeping the boundary fixed. The state that remains is guaranteed to have only extended bonds internally; we then reset the \emph{initial} lengths $\ell_{0,k}$ to their relaxed values, after which we can use the procedure detailed around Eq.~(\ref{eq:setelleq}) to set the self-stress $\Pi$ by varying the \emph{equilibrium} lengths $\ell_\mathrm{eq}$. We then deform our frames by imposing a displacement to the four corners. This deformation is prescribed by the deformation gradient tensor $\Lambda$, which is generally parametrized by some scalar measure of the strain. We then let the positions of all non-corner nodes relax (which amounts to exploring the space of non-affine node displacements $\vec \delta_i$) until a minimum of the energy Eq.~(\ref{NAEn}) is reached. Then, we expand this minimized energy $\calE(\Pi)$ around its minimum for given $\Lambda$ and $\Pi$ to second order in the strain variable. Because of the minimization, the linear term in this expansion is zero; the modulus is the coefficient of the quadratic term.

We will mostly be considering the case of simple shear, for which the boundary points are displaced according to the deformation gradient tensor
$$\Lambda(\gamma) = \begin{bmatrix}
    1 & \gamma \\
    0 & 1
  \end{bmatrix}\, .
$$
In this case, the scalar strain measure is the shear strain $\gamma$. Our main finding is that a nonzero value of $\Pi$ does, indeed, augment the shear modulus $\mu$. Fig.~\ref{fig:affineMod} illustrates this point. The linear shear modulus is seen to vary with the self-stress over the range of physically relevant self-stresses (i.e., those corresponding to positive equilibrium lengths). Throughout this regime, two behaviors stand out: Firstly, in the mechanically stable regime ($\mu>0$), the shear modulus changes with $\Pi$, and secondly,  self-stress can induce mechanically unstable ($\mu<0$) states. This response is generic; although the values of $\Pi_{\rm min}$ and $\Pi_{\rm max}$ differ from frame to frame, we see the same behavior in all frames we have analyzed. This then is finding ({\em i}): as opposed to affinely deforming frames where there can be no effect, {\em }self-stress controls the linear mechanical response in generic, non-affinely deforming frames}. 

Digging down into the nature of this dependency, we focus first on linear response in the mechanically stable regime. The mechanism by which self-stress controls the linear modulus is laid out in Section \ref{Sec2}: the set of $\{\vec \delta_i\}_{i=1}^{N_b}$ at which the minimum of energy Eq.~(\ref{NAEn}) is attained given a deformation $\Lambda(\gamma)$ depends, through the coupling term Eq.~(\ref{NAcoup}), on the pre-stress. 

As detailed in Sec.~\ref{Sec2}, the shear modulus of a non-affinely deforming \emph{pre-stressed} frame can never exceed the shear modulus of the same frame deforming affinely, which in turn is equal to the shear modulus of the \emph{non-prestressed} frame deforming affinely. We thererfore have $\mu\leq\mu_\mathrm{aff}$.

\begin{figure}
\includegraphics[width=\columnwidth]{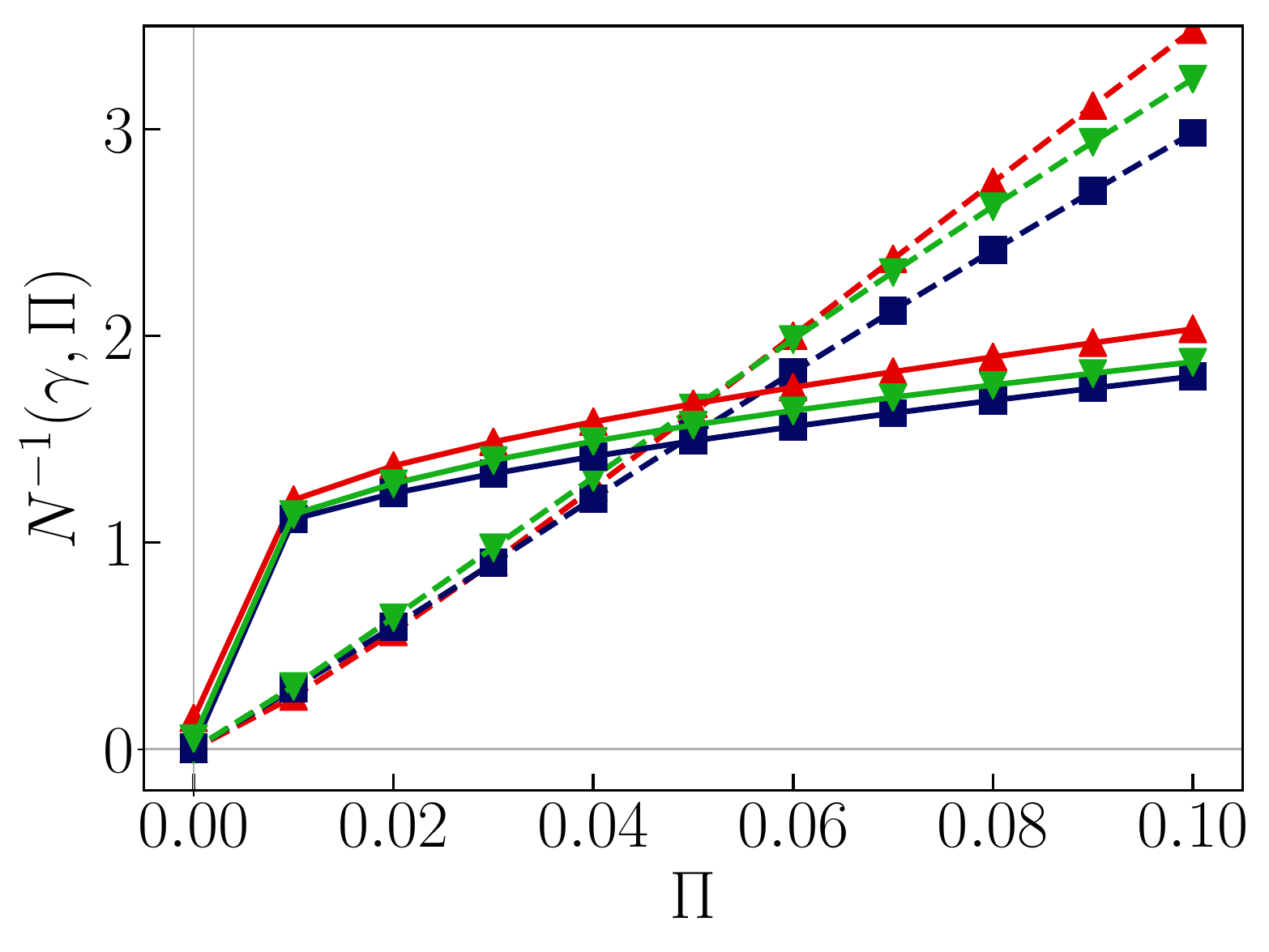}
\caption{~Pre-stress dependence of $N^{-1}(\gamma, \Pi)$ for non-affinely deforming self-standing frames, averaged over an ensemble of 10 network realizations each for 40 nodes (red $\filledmedtriangleup$), 50 nodes (dark blue $\filledmedsquare$) and 60 nodes (green $\filledmedtriangledown$). As the self-stress $\Pi$ is increased the non-affinity drops, approaching zero corresponding to the affine limit. The closeness of the curves emphasizes that the system size dependence is small. Dashed lines represent the case of simple shear, for which $N\sim 1/\Pi$ appears to describe the data well over this range of self-stress values (although it should be noted that $N(0)$ is actually finite). Solid lines describe the non-affinity for pure shear. Uniform compression shows similar behavior as pure shear (not shown).}
\label{fig:nonaffinity}
\end{figure}

From Fig.~{\ref{fig:affineMod}}, it appears that for higher pre-stresses the shear modulus approaches its affine limit. This could happen for a variety of reasons, but the most obvious one is that the non-affine displacements of the nodes $\{\vec \delta_i\}_{i=1}^{N_s}$ themselves tend to zero. In Sec. \ref{Sec2}, we demonstrate this to be the predicted dependence for frames possessing only tensile self-stresses on the non-boundary bonds. To verify this prediction, we measure the non-affinity, defined as
\be
N(\gamma,\Pi)\equiv\frac{1}{\gamma^2}\left\langle\left(\frac{|\vec \Delta_{k}|}{\ell_{0,k}}\right)^2 \right\rangle\, ,
\ee
with the average $\langle\cdot\rangle$ running over all $N_b$ springs. Because $N$ becomes very large as $\Pi\to0$, we plot $1/N$ in Fig.~(\ref{fig:nonaffinity}) for both simple shear and pure shear deformations. Indeed, increasing $\Pi$, but keeping it below its maximal value, suppresses the magnitude of the non-affine displacements. By extension, the shear modulus should approach its affine limit.

\begin{figure}[!tb]
\includegraphics[width=\linewidth]{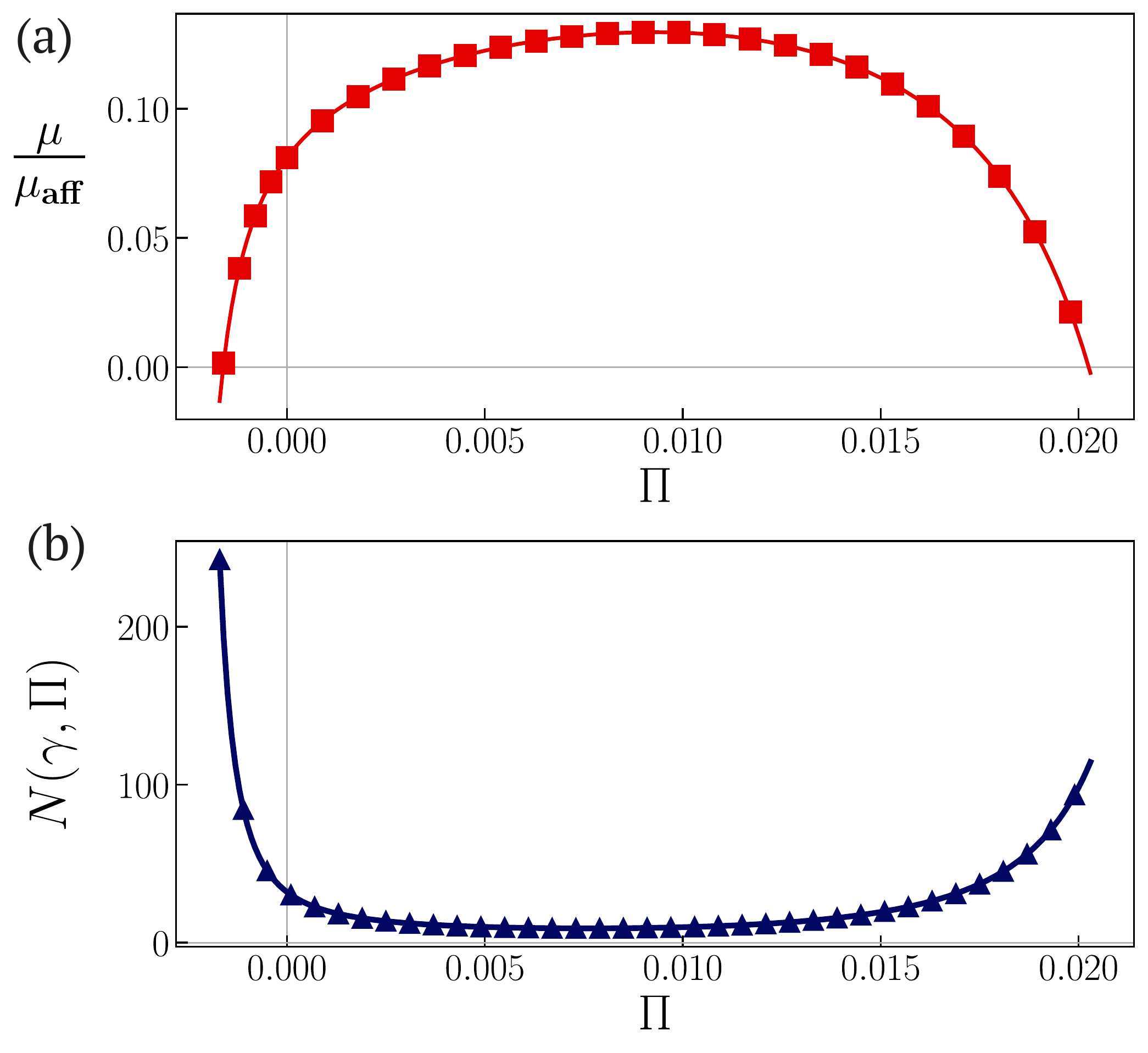}
\caption{(a) Pre-stress dependence of the simple shear modulus of a frame whose SSS contains both tensile and compressive bonds. In such frames, the range of pre-stresses compatible with mechanical stability is bounded both from below and above. (b) As in the purely tensile frames, the the non-affinity measure signals the instability, blowing up at the stability edges.}
\label{mixNA}
\end{figure}

The non-affinity measure $N(\gamma, \Pi)$ signals other interesting behaviors. First, it becomes very large (albeit finite, as it should be in a finite-size system) as the system approaches a mechanical instability, marked by the shear modulus becoming negative. In Section \ref{Sec4} we investigate this regime further, for now we note that apparently large deviations from affine occur at such a point; these correspond to the system acquiring novel state(s) of mechanical equilibrium, possibly quite far away from the reference configuration. This finding echoes the signalling quality of the non-affinity also observed in \cite{ChaseIsostatic}. There, too, a cusp in a quantity directly proportional to our $N(\gamma, \Pi)$ marks the floppy-to-rigid transitions occurring in both flexibly hinged and bond-bending fiber networks with periodic boundary conditions. 

In case all internal forces have the same sign, we find that the frame becomes unstable if we make all internal bonds carry compressive loads, in accordance  with the known destabilizing nature of compressive forces~\cite{Alexander,WyartPRE2005,VanHecke2010a,LiuNagelReview}. Furthermore, our finding that the deformation of the frame becomes increasingly affine as we ramp up the self-stress confirms earlier reports, that---for various different systems both in 2D and 3D---tensile pre-stresses increase affinity~\cite{CioroianuJMMPS2016,WenBasuSM2012, huisman2011internal}. We do note, however, that conflicting reports also exist which appear to show the non-affinity rising with pre-stress. Based on what we are able to conclude here, we cannot rule out that certain systems are more compression-dominated in which case the interdependence between non-affinity and pre-stress could well be oppositely signed.

The fact that tensile pre-stresses suppress non-affinity is not unique to shear deformation. When we subject our frames to pure shear, effected by using
$$
\Lambda(\gamma_P) =  \begin{bmatrix}
    1+\gamma_P & 0 \\
    0 & 1-\gamma_P
  \end{bmatrix}~,
$$
and uniform compression, for which ($\epsilon>0$)
$$
\Lambda(\epsilon) =  \begin{bmatrix}
    1-\epsilon & 0 \\
    0 & 1-\epsilon
  \end{bmatrix}~,
$$
we see the same behavior. Fig.~\ref{fig:nonaffinity} shows this for simple and pure shear;  in each of these cases the non-affinity drops as $\Pi$ is increased, indicating that for general deformations and in frames with purely tensile SSS's, large positive self-stresses suppress non-affinity and steer the associated moduli to their affine values. Interestingly, this effect appears strongest (in the sense, that the range of pre-stresses compatible with stability is smallest) for simple shear, but regardless: Since any deformation in 2D may be decomposed into shear and extensional components this establishes finding ({\em ii}): a coupling between the pre-stress and the non-affinity controls the $\Pi$-dependence of $\mu$. Fig.~\ref{mixNA} shows, that this coupling is also manifested in frames with a mixed SSS (that is, a SSS in which some bonds are compressed and others are tensed) and that, completely analogous to what happens in purely tensile frames, a steeply increasing non-affinity foreshadows mechanical instability.

\section{Nonlinear response of self-stressed frames: Multistability}\label{Sec4}
\begin{figure}[!tb]
\includegraphics[width=\linewidth]{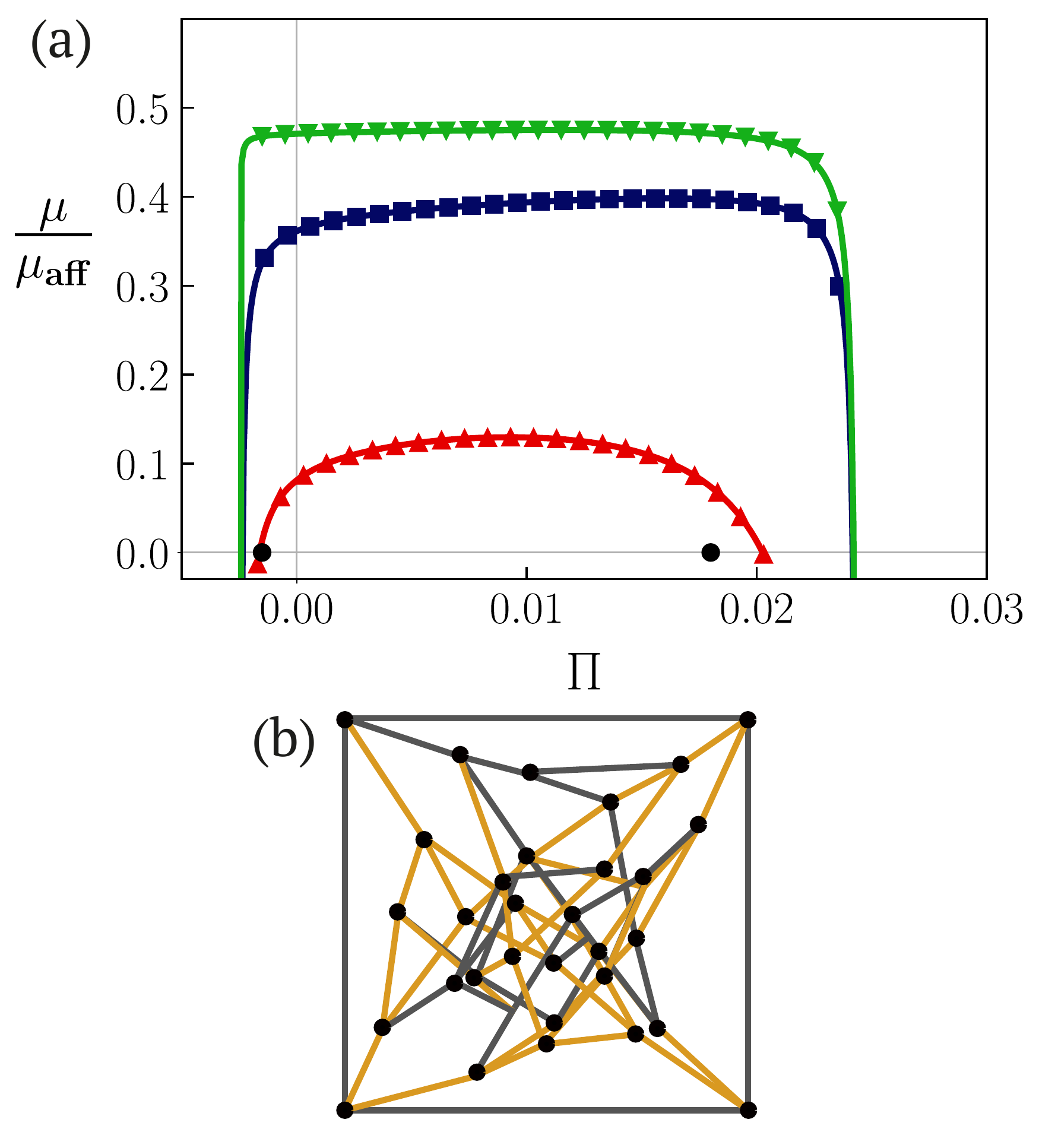}
\caption{(a)~The simple shear (red $\filledmedtriangleup$), bulk (green $\filledmedtriangledown$) and pure shear (dark blue $\filledmedsquare$) modulus (normalized by the affine value $\mu_\mathrm{aff}$) are shown for  $-0.0025 < \Pi < 0.025$. (b)~The corresponding frame has both tensile and compressed bonds, so that both positive and negative $\Pi$ will drive the system unstable. The black circles indicate the values of $\Pi$ at which the smallest eigenvalue of the Hessian becomes negative.}
\label{fig:unstableMod}
\end{figure}
We now turn to the behavior of frames whose reference geometry has been rendered unstable by sufficiently large (or small) pre-stress. Their instability is marked by a negative linear modulus. It is important to note that these instabilities generally do {\em not} occur at values of the self-stress corresponding to $\Pi_{\rm min}$ or $\Pi_{\rm max}$. We will return to how the instability thresholds may be computed in a moment. 

For an unstable network, a negative shear modulus means that the expansion of the energy in the strain variable (which we will assume to be the shear strain $\gamma$ for now) starts out as
\be
\calE(\gamma,\Pi)\approx \calE(0,\Pi)-\frac12 |\mu(\Pi)|\gamma^2+{\cal O}(\gamma^3)\,.
\ee
This energy will prompt a spontaneous shear strain $\gamma$, adjusting the shape of the box (and, as this changes, the positions of the nodes) until a new, stable geometry is reached. Because
the frame at $\gamma=0$ is in mechanical equilibrium (albeit an unstable equilibrium), there is no linear term in $\gamma$. Generally, networks generated using our Henneberg construction protocol have both tensile and compressive elements in the single state of self-stress spanning the network. As we have seen in the previous paragraph, this means the original square shape of the frame will become unstable if the pre-stress becomes too high in either direction. 

Identifying these instabilities via specific moduli gives the values of pre-stress for which deformation along the one-parameter axis corresponding to that modulus becomes unstable. Within the network, these instabilities can be seen earlier by looking at the the lowest eigenvalue of the Hessian matrix of the total spring energy, which becomes negative as soon as one mode becomes unstable. Note that a single unstable mode is not guaranteed to lead to instabilities in the moduli, as the displacements of the boundary points according to that mode generally do not match the imposed deformation gradient tensor $\Lambda$.

As is seen in Fig.~\ref{fig:unstableMod}, we find that destabilizing the system tends to happen at the lowest pre-stress for simple shear; to explore the consequences of destabilization we continue to focus on simple shear from here onwards. 

\begin{figure}[!tb]
\includegraphics[width=\linewidth]{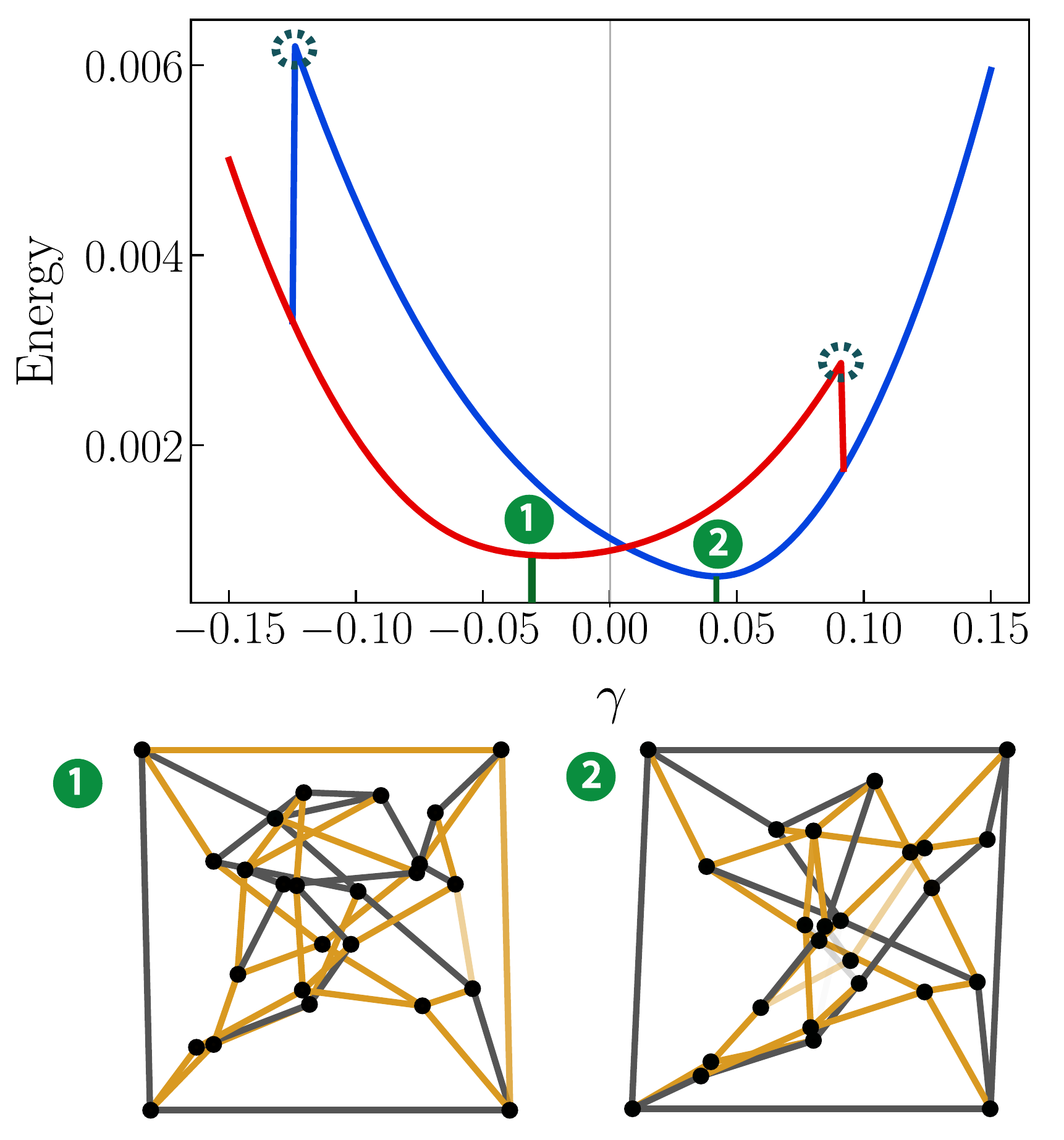}
\caption{Energy as a function of shear strain for a bistable frame with hysteresis. Only the two branches of the limit cycle are shown, with the points at which the frame jumps to the lower branch indicated. The frame geometries at the bottom correspond to the two minima labeled by (1) and (2).}
\label{fig:twobranches}
\end{figure}

\begin{figure}[!tb]
\centering
\includegraphics[width=\linewidth]{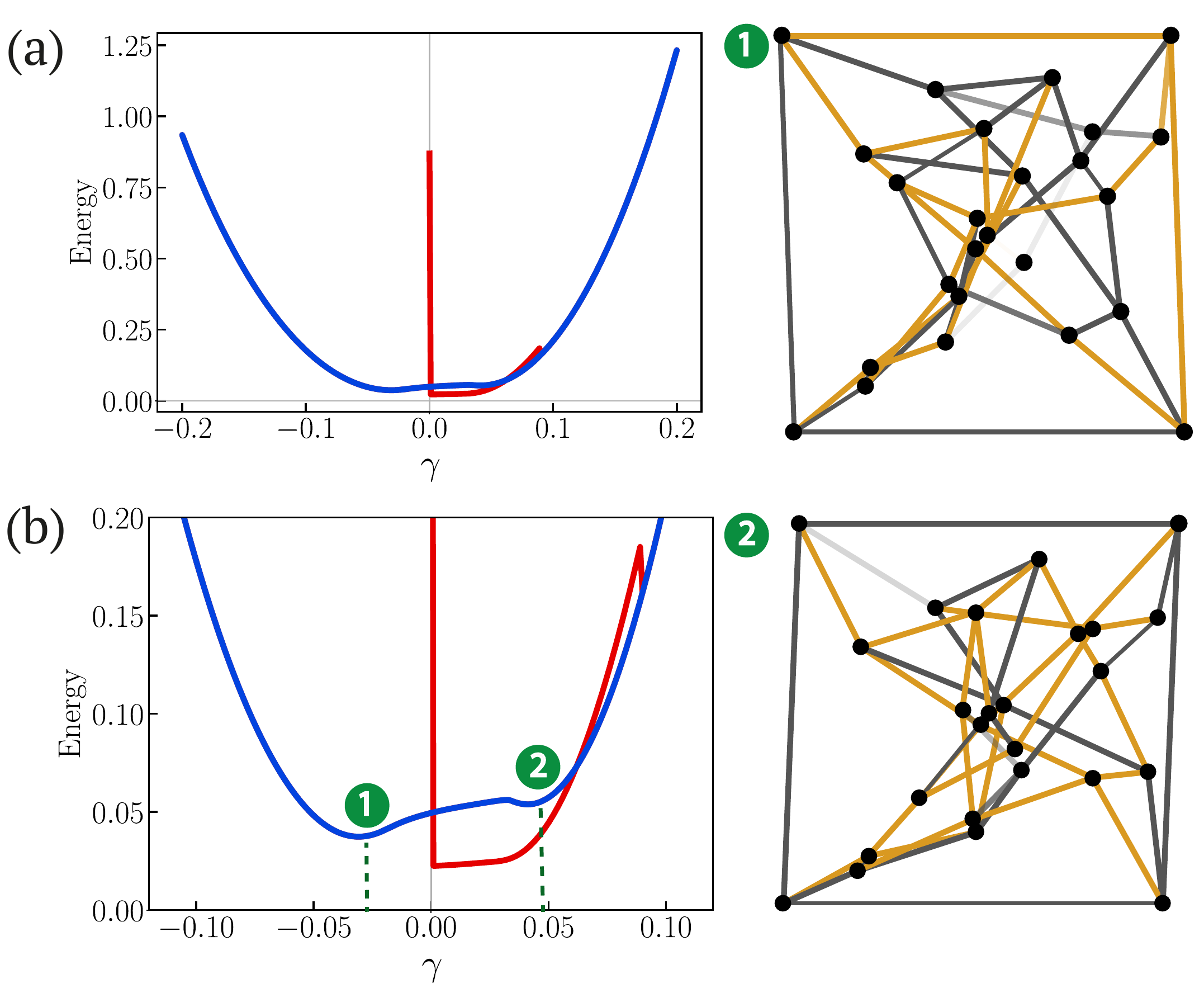}
\caption{(a) Energy versus shear strain $\gamma$ for $\Pi = 0.009$. The first shear cycle (red) is marked with an immediate drop in the energy, followed by small jump to a different branch at roughly 10\% strain. The rest of the first and second (blue) shear cycle continues on this branch, which displays two local energy minima at $\gamma=-0.036$ and $\gamma=0.041$, as illustrated in the zoom-in graph (b).  The geometries of the frame in these local energy minima are shown on the right. The tensile bonds are shown in gray while the compressed bonds are in yellow and the opacity of the bond is proportional to the tension $t_\mathrm{ss,k}$ of each bond.}
\label{fig:bistable}
\end{figure}

\begin{figure}[!tb]
\centering
\centering
\includegraphics[width=\linewidth]{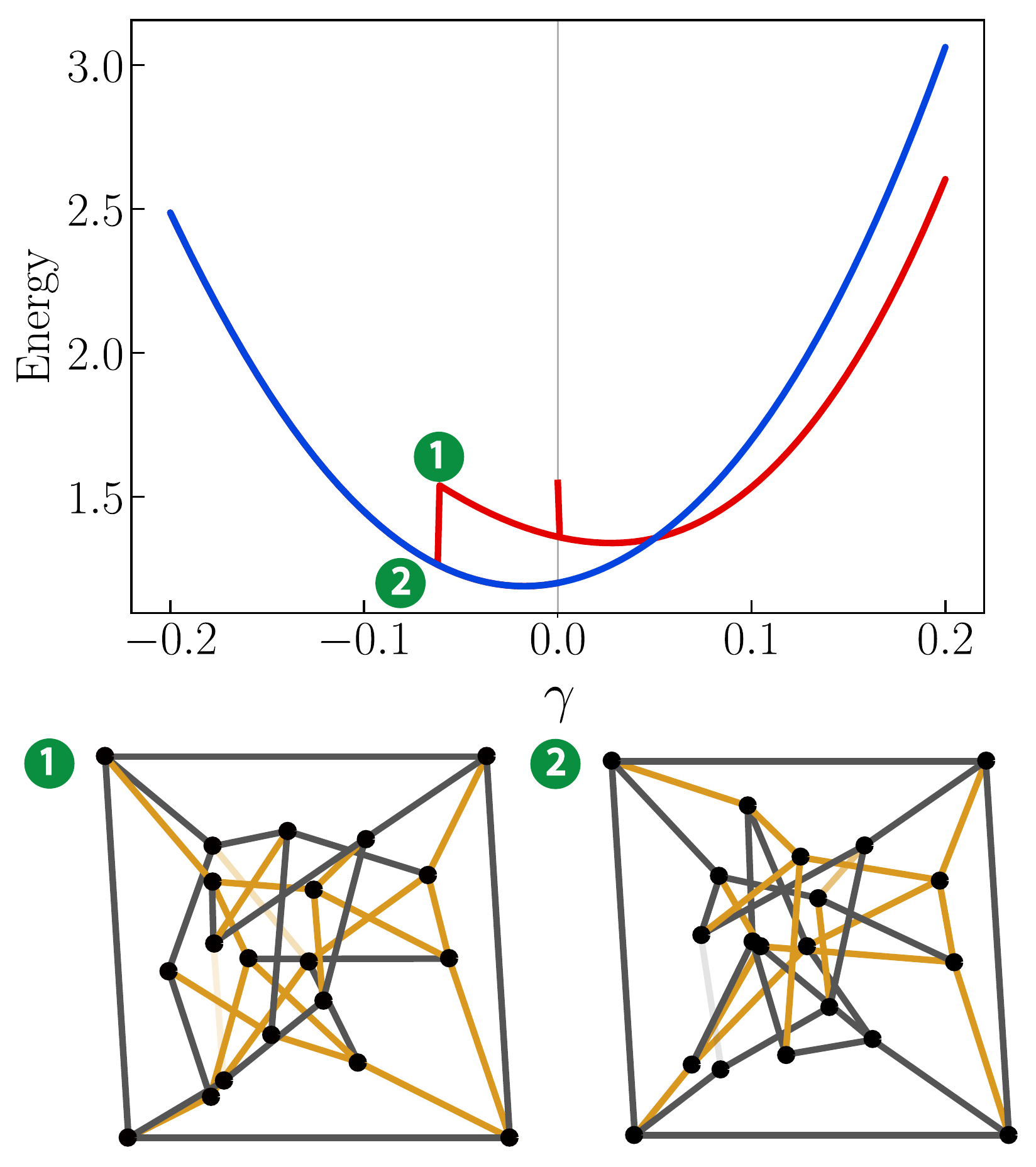}
\caption{ The energy of the shear deformation is plotted at every strain step $\gamma$ plotted for each cycle. The first cycle is represented in red and second cycle is represented in blue. Note that the following shear cycles follow the blue curve, hence is not plotted. At $\Pi = 0.02$ and $\gamma=-0.062$ the frame makes a hysteretic jump to the blue branch and continues to follow that path for all the negative values of $\gamma$. Interestingly, for $\gamma>0$ the blue curve has greater curvature than the red cycle, indicating an increase in the shear modulus. The local distribution of stresses in the frame is highly heterogeneous, and the opacity of the bonds is scaled to the amount of stresses borne by them. }
\label{fig:rerrange_mono}
\end{figure}

Once the system has become unstable to simple shear, allowing the system to find its preferred shape along the shear deformation direction should then reveal new stable states that have a non-square shape. We have studied the corresponding energy landscape as a function of shear strain $\gamma$ and found that in many cases there was more than one non-square stable shape. Thus, our model system  can be tuned from monostable (and square) to bistable (switchable between two different shapes) by varying the internal pre-stress, while keeping the initial geometry identical.

We find two distinct types of bistability. Most common is the emergence of two branches in the energy function, with discontinuous jumps between them and hysteresis over the course of a strain cycle, as seen in Fig.~\ref{fig:twobranches}. In rare cases, we also see unstable behavior where the energy remains smooth as a function of strain, and where the initially stable state at zero shear is destabilized at finite $\Pi$ giving way to two local minima (see Fig.~\ref{fig:bistable}).

The self-stressed frames have a positive shear modulus at $\Pi = 0$, as expected since the elastic energy $\mathcal{E}$ is non negative and the energy of the reference configuration is $\mathcal{E}(0,0)=0$ (plus our frames have no floppy modes). The frames are subjected to multiple cycles of simple shear until their energy curve attains a steady state. The energy curve is calculated through numerical minimization. There is no noticeable unstable behavior at $\Pi = 0$, even though some frames can change their energy path after the first cycle, but the change is smooth. Multistability is most pronounced at nonzero values of pre-stress.

One such instability is noticed when the pre-stress is set to values at which the linear shear modulus is negative. At these values, the energy drops instantly at the first shear step, prompting rearrangement of the network, as can be seen in Fig.~\ref{fig:bistable}a and Fig.~\ref{fig:rerrange_mono}. 
Fig.~\ref{fig:twobranches} also shows that as these frames are sheared, they undergo snap changes in the geometry at strains where the energy jumps between branches corresponding to different shear cycles. The change in distribution of stresses at each strain step makes such hysteretic jumps between energy branches very common in these finite frames. The drop in energy is typically paired with a significant spatial rearrangement of the nodes and a decrease in the number of compressed bonds. In some networks we observed a phenomenon reminiscent of strain hardening: they showed a larger shear modulus in the second shear cycle than in the first. 

The hysteretic self-stressed frames thus constitute a mechanical memory device: The overall shape of the frame is indicative of the direction in which the frame was sheared last, and in some cases (in addition to looking at the shape) this information can also be read out by probing the modulus. A closely related behavior is shown in Fig.~\ref{fig:rerrange_mono}: This frame is unstable in its initial configuration, snaps immediately to a metastable branch, but ultimately goes through another instability after which it becomes fully stabilized in a new, non-square shape. This particular frame can hence report whether or not it has ever been sheared to the left by more than a certain amount (in this example: 7\%), serving as another type of mechanical sensor.

\section{Discussion}
We have studied the effects of self-stresses on the linear and nonlinear mechanical properties of freestanding spring networks or \emph{frames}. By keeping the geometry constant and adjusting the self-stress via the rest lengths of the springs, we isolated the direct effects of the stresses from those mediated by the changes in frame geometry. Within the scope of our model, the energy involved in affine deformations of the frames is independent of the self-stress, so that the moduli can only change if the deformations are non-affine. We have shown analytically and numerically that in frames in which the internal bonds are all tensile, the self-stress suppresses non-affinity.

In the context of tensegrities or free-standing frames, it was already known that engaging the self-stress cannot lift a floppy mode to become rigid~\cite{schenk2007zero}. This result does not apply to our frames, since these, by construction, have a nonzero shear modulus at $\Pi=0$. Comparing this to our general finding, neither result is more general than the other: Our result, while being applicable to frames that are already rigid at zero self-stress, does rely on the interactions being harmonic springs with a stiffness that is inversely proportional to their equilibrium length.

The extent to which frames \emph{do} deform affinely is influenced by many factors. First, there may be symmetries in the system that dictate affine deformations. Second, it has been known for decades that networks of springs with zero rest length (such as Gaussian chain networks, an idealized model for rubbers), deform affinely regardless of the pre-stress and the distribution of stiffnesses~\cite{James1947}. Third, we know that networks that are more densely connected, with many more bonds per node than the isostatic condition demands, typically deform more affinely~\cite{wyart08,ChaseIsostatic,ellenbroek2009jammed}. And finally, even systems that do not satisfy any of the above criteria can be made to deform more affinely by way of tensile pre-stresses. This feature---one of the central findings of this work--- is in line with the general stabilizing tendency of tensile pre-stress noted in \cite{Alexander}. While the effect was observed in simulations of a unit cell model for polymer networks~\cite{CioroianuJMMPS2016}, it does not appear to have been explicitly demonstrated, verified or derived for many other systems, although many studies on non-affinity in disordered networks are, retrospectively, consistent with our findings ~\cite{WenBasuSM2012,Onck2005,huisman2011internal,ChaseIsostatic}. Our findings suggest that it is not strain itself, but rather the tensions incurred as a system is strained that suppresses non-affinity in these earlier works. Self-stressed frames---tensed but otherwise undeformed systems---allow us to separate the two, proving our point.

Here, we showed that frames with compressive or mixed SSS may become unstable upon increasing the self-stress, and that the approach of the instability is marked by both a decreasing modulus and increasing (possibly diverging) non-affinity, while purely tensile self-stresses generally suppress non-affinity.

Going beyond linear properties, we showed that the destabilizing nature of self-stresses leads to frames with interesting energy-strain profiles. These feature strain-induced rearrangements, corresponding to jumps between branches. Some of these show multiple stable minima, with hysteretic behavior between them; others are initially metastable and snap to a single stable minimum after a sufficiently large shear deformation. We speculate that such hysteretic cycles can be used to design self-stress controlled mechanical switches that remember (store) the which direction in which they were last sheared. Likewise, The metastable case can serve as a sensor; reporting whether an object has previously been deformed by more than a certain amount. This brings new uses for self-stress-controlled nonlinear mechanical properties to mechanical metamaterials~\cite{florijn2014programmable}.

The use of self-stresses to enhance the mechanical properties of network materials may well be broader than captured by freestanding frames. Our results suggest, that interesting effects are likely to arise whenever not all self-stresses are tensile. We speculate that similar effects can play a role at the microscopic scale, enhancing the already spectacular mechanical properties of double and triple network elastomers~\cite{Ducrot2014}. An effective way to harness the benefits of self-stresses would be to introduce elements into these materials that are better able to carry compressive loads, such as semiflexible polymers~\cite{bai11}, or to explore strategies that build up compressive loads automatically~\cite{Fern2018}.   

Unlike the frames we study here, larger systems will likely have many states of self-stress. Preliminary calculations have shown that the second and subsequent SSS(s) have a decreasing contribution to the mechanical response of the network. More work is required to assess the differences between localized states of self-stress, which may be used for patterning materials, and those that are system-spanning, and likely have a more direct effect on the macroscopic mechanical properties. 

In summary, and returning to issue {\em(iii)} as defined in the introduction, we have shown that mechanical preconditioning, by means of states of self stress, represent a powerful control modality able to tune the linear and nonlinear mechanical response of disordered network materials. Designer matter, both macroscopic and molecular, has only just begun to exploit this design principle, which we suspect may be far more broadly leveraged to create new adaptive materials with tailored response.



\section{Acknowledgments}
We are grateful to Thijs van der Heijden for technical advice. This work is part of the research programme on Computational Sciences for Energy Research (14CSER005), which is financed by the Netherlands Organization for Scientific Research (NWO) and Royal Dutch Shell, and of the research programme on Marginal Soft Matter (FOM12CSM01), which is also financed by NWO.

\appendix
\section{Affine deformation with SSS}
\label{sec:Appendix}
In this appendix, we demonstrate that, with our choice of spring constants
$k_i=\frac{Y_i}{\ell_{eq,i}}$, and in free-standing frames, the effect of
engaging the state of self-stress on the total energy of an affine deformation is independent of strain,
that is
$$\mathcal{E}(\Pi) - \mathcal{E}(0) = \mathcal{C}(\Pi)~.$$
The proof amounts to showing that both terms in Eq.~(\ref{eq:ediffmultiline})
that depend on the displacements of the nodes are proportional to the overall
stress in the frame, which is zero in free-standing frames since there are no
external forces acting on them. In general, the Cauchy stress tensor for these frames is
\begin{equation}
\label{eq:stresstensordef}
\sigma_{\alpha\beta}= \frac{\Pi}{V}\sum_{k=1}^{N_b}\ell_{0,k}  {t}_{ss,k}\hat n_{k,\beta}\hat n_{k,\alpha}\,
\end{equation}
with $V$ the system volume. By definition it is zero for all values of $\Pi$, which means the sum itself must equal zero. Note that nothing below assumes that the displacements or the strains are small, so the result
will be valid for nonlinear deformations.

The change in the length of any bond: $\delta l = \ell - \ell_0$ can be written
as the displacement of the nodes that it joins. $\vec{u}_{ij}$ is the
displacement vector and $\vec{r}_i$, $\vec{r}_j$ are the position vectors to
the $i^{th}$ and $j^{th}$ node such that: $\ell_0 \, \hat{n}=\vec{r}_j -
\vec{r}_i$
\begin{align}
\vec{u}_{ij} &= \Lambda\cdot(\vec{r}_j - \vec{r}_i) - \mathbb{1}_2(\vec{r}_j - \vec{r}_i) \nonumber \\
&=(\Lambda - \mathbb{1}_2)\cdot(\vec{r}_j - \vec{r}_i)
\end{align}
where $\Lambda$ is the displacement gradient tensor. The fact that we are considering affine deformations corresponds to
the application of this global tensor at the level of individual bonds. Using Einstein notation,
\begin{align}
u_\parallel &= \hat{n}\cdot\vec{u}_{ij}\\
&=\hat{n}_\alpha\cdot(\vec{u}_{ij})_\beta\\ \nonumber
&=\ell_0\,(\Lambda - \mathbb{1}_2)_{\alpha \beta}\, \hat{n}_\beta\,\hat{n}_\alpha
\end{align}
Thus, we can write the linear term (in $u_\parallel$) in Eq.~(\ref{eq:ediffmultiline}) as
\begin{eqnarray}
\sum_{k=1}^{N_b} t_{ss,k} \, u_{k,\parallel} &=& (\Lambda -\mathbb{1}_2)_{\alpha \beta} \sum_{k=1}^{N_b}\ell_{0,k}  {t}_{ss,k}\hat n_{k,\beta}\hat n_{k,\alpha}\nonumber\\
&\sim&(\Lambda -\mathbb{1}_2)_{\alpha \beta}\sigma_{\alpha\beta}~,
\end{eqnarray}
which is zero in free-standing frames.

Secondly, consider the identity
\begin{align}\label{perpparallel}
u_\parallel^2 + u_\perp^2 &= \vec{u}_{ij}\cdot\vec{u}_{ij} \\ &= \ell_0\,(\Lambda-\mathbb{1}_2)_{\mu \alpha}\hat{n}_{k,\alpha}\, \ell_0(\Lambda - \mathbb{1}_2)_{\mu \beta}\hat{n}_{k,\beta}~. \nonumber
\end{align}
Using this identity, the quadratic term in Eq.~(\ref{eq:ediffmultiline}) can be written as
\begin{align}
& \sum_{k=1}^{N_b}\left(\frac{{t}_{ss,k}}{\ell_{0,k}}\right)(u_{k,\parallel}^2 + \, u_{k,\perp}^2) \\ 
&= \sum_{k=1}^{N_b} \left(\frac{t_{ss,k}}{\ell_{0,k}}\right) \ell_0(\Lambda-\mathbb{1}_2)_{\mu \alpha}\hat{n}_{k,\alpha}\, \ell_0(\Lambda - \mathbb{1}_2)_{\mu \beta}\hat{n}_{k,\beta} \nonumber \\
&= (\Lambda-\mathbb{1}_2)_{\mu \alpha}(\Lambda - \mathbb{1}_2)_{\mu \beta}\sum_{k=1}^{N_b}\ell_{0,k}t_{ss,k}\hat{n}_{k,\alpha}\hat{n}_{k,\beta}\nonumber \\
&\sim  (\Lambda-\mathbb{1}_2)_{\mu \alpha}(\Lambda - \mathbb{1}_2)_{\mu \beta}\sigma_{\alpha\beta}~.
\end{align}
Again, this term is zero for free-standing frames, completing the proof.

\end{document}